\documentstyle[11pt]{article}

\newcommand{\vn}{V N N}

\newcommand{\ra}{\rightarrow}
\newcommand{\aleq}{\stackrel{<}{\sim}}
\newcommand{\ageq}{\stackrel{>}{\sim}}
\newcommand{\gtuv}{\tilde{g}^{\mu\nu}}
\newcommand{\gtab}{\tilde{g}^{\alpha\beta}}
\newcommand{\ktu}{\tilde{k}^{\mu}}
\newcommand{\ktv}{\tilde{k}^{\nu}}
\newcommand{\Ord}{O\left( \frac{1}{p_{L}^{2}} \right)}
\newcommand{\nn}{\nonumber}
\newcommand{\bg}{\begin{eqnarray}}
\newcommand{\en}{\end{eqnarray}}

\begin{document}
%%%%%%%%%%%%%%%%%%%%%%%%%%%%%%%%%%%%%%%%%%%%%%%%%%%%%%%%%%%%%%%%%%%%%%%%%%%%%%%
\vspace*{-2cm}

\large

\vspace*{0.4cm}

\huge
\begin{center}
Role of Vector Mesons in High-$Q^{2}$ Lepton-Nucleon Scattering.
\end{center}

\vspace*{0.4cm}

\large
\begin{center}
W.Melnitchouk and A.W.Thomas                            \\
Department of Physics and Mathematical Physics          \\
University of Adelaide                                  \\
Box 498 G.P.O., Adelaide, 5001, Australia
\end{center}

\vspace*{0.2cm}
%%%%%%%%%%%%%%%%%%%%%%%%%%%%%%%%%%%%%%%%%%%%%%%%%%%%%%%%%%%%%%%%%%%%%%%%%%%%%%%
\normalsize
\begin{center}
Abstract
\end{center}

\vspace*{0.2cm}

\hspace*{-0.5cm}
The possible role played by vector mesons in inclusive deep inelastic
lepton-nucleon scattering is investigated.
In the context of the convolution model, we calculate self-consistently
the scaling contribution to the nucleon structure function using the
formalism of time-ordered perturbation theory in the infinite momentum frame.
Our results indicate potentially significant effects only when the
vector meson---nucleon form factor is very hard.
Agreement with the experimental antiquark distributions, however,
requires relatively soft form factors for the $\pi N$, $\rho N$ and
$\omega N$ vertices.

\vspace*{6cm}
To appear in Phys.Rev.D

\newpage

%%%%%%%%%%%%%%%%%%%%%%%%%%%%%%%%%%%%%%%%%%%%%%%%%%%%%%%%%%%%%%%%%%%%%%%%%%%%%%%
\section{Introduction}

In the context of meson exchange models of the $N$ force in nuclear physics,
it has long been realised that vector mesons play a vital role
\cite{potl,bonn}.
For example, the isovector $\rho$ meson is needed to provide sufficient
cancellation of the tensor force generated by $\pi$ meson exchange, which
would otherwise be too large.
On the other hand, the isoscalar $\omega$ meson, through its large vector
coupling, is responsible for the short range $NN$ repulsive force, and also
provides most of the spin-orbit interaction.
Traditionally it has been necessary to use hard vector meson---nucleon
form factors in order to fit the $NN$ phase shifts \cite{bonn}.
However, alternative approaches have recently been developed in which the
$NN$ data can be fitted with quite soft form factors \cite{boch,thol}.

{}From another direction, the vector meson dominance model of the elastic
electromagnetic nucleon form factors, in which an isovector photon
couples to the nucleon via a $\rho$ meson, provides a natural explanation
of the dipole $Q^{2}$ behaviour of the $\gamma N N$ vertex function.
Recent analyses \cite{boch} have shown that a $\rho N N$ vertex parameterised
by a soft monopole form factor ($\Lambda_{1} \sim 800$ MeV) provides a good
description of the $Q^{2}$ dependence of the Dirac and Pauli form factors.
The effect of vector mesons upon nucleon electromagnetic form factors has
also been explored \cite{jc} in the cloudy bag model \cite{cbm},
and in various soliton models \cite{sol}.

In this paper we investigate the possible role played by vector mesons
in high-$Q^{2}$ inelastic inclusive scattering of leptons from nucleons,
in the context of the so-called convolution model, in which the deep inelastic
process is described in terms of both quark and explicit meson-baryon
degrees of freedom.
More specifically, the scaling property of the meson- and baryon-exchange
contributions to the inclusive cross section allows us to probe
the extended mesonic structure of nucleons.

Quite naturally the pion, being by far the lightest meson, was the first
meson whose contributions to the nucleon structure function were investigated
\cite{first}.
It was later noticed \cite{t83} that the pion cloud could be responsible for
generating an asymmetry between the $\bar{u}$ and $\bar{d}$ quark content
of the proton sea, through the preferred proton dissociation into a neutron
and $\pi^{+}$.
Furthermore, deep inelastic scattering (DIS) data on the momentum fractions
carried by antiquarks were used to obtain an upper limit on this
non-perturbative pionic component \cite{t83,fms}.
An enhancement of $\bar{d}$ over $\bar{u}$ resulting from this process was
also postulated as one explanation for the slope of the rapidity
distribution in $p$--nucleus Drell-Yan production \cite{ito}.
More recently it has been hypothesised that this asymmetry could account for
some of the apparent discrepancy between the naive parton model prediction for
the Gottfried sum rule \cite{got} and its recently determined experimental
value \cite{nmc}, and indeed this has resulted in the greater attention that
the convolution model of lepton-nucleon scattering has received [14-21].

In a model in which the nucleon has internal meson and baryon degrees of
freedom, the physical nucleon state in an infinite momentum frame can be
expanded (in the one-meson approximation) in a series involving bare
nucleon and two-particle meson---baryon states:
\bg
\left| N \rangle_{\rm phys} \right.
& = & \sqrt{Z}\
\left\{ \left| N \rangle_{\rm bare} \right.
\ +\ \sum_{M B} \int dy\ d^{2}{\bf k}_{T}\
     g_{0_{MBN}}\ \phi_{MB}(y,{\bf k}_{T})\
     \left| M (y,{\bf k}_{T}); B (1-y,-{\bf k}_{T}) \rangle \right.
\right\}.
\label{state}
\en
Here, $\phi_{MB}(y,{\bf k}_{T})$ is the probability amplitude for the physical
nucleon to be in a state consisting of a meson $M$ and baryon $B$,
having transverse momenta ${\bf k}_{T}$ and $-{\bf k}_{T}$, and
carrying longitudinal momentum fractions $y$ and $1-y$, respectively.
$Z$ is the bare nucleon probability.
Although we shall work in the one-meson approximation, we will include
higher order vertex corrections to the bare coupling constants $g_{0_{MBN}}$.
Illustrated in fig.1 is the deep inelastic scattering of the virtual photon
from the two-particle state $| M; B \rangle$.
In fig.1a the photon interacts with a quark or antiquark inside the exchanged
meson, while in fig.1b the scattering is from a quark in the baryon
component of the physical nucleon.

According to eqn.(\ref{state}), the probability to find a meson inside a
nucleon with momentum fraction $y\ (= k \cdot q / p \cdot q)$\ is
(to leading order in the coupling constant)\
$f_{MB}(y) \equiv Z\ g_{0_{MBN}}^{2}\
                  \int d^{2}{\bf k}_{T}\
                  \left| \phi_{MB}(y,{\bf k}_{T}) \right|^{2}$.
This must also be the probability to find a baryon inside a nucleon with
momentum fraction $1-y$.
The baryon distribution function, $f_{BM}(y')$, where
$y' = p' \cdot q / p \cdot q$, is probed directly through the process in
fig.1b, and should be related to the meson distribution function by
\bg
f_{MB}(y) & = & f_{BM}(1-y)
\label{pcon}
\en
for all $y$, if the above interpretation is valid.
We also demand equal numbers of mesons emitted by the nucleon,
$\langle n \rangle_{MB} = \int_{0}^{1} dy\ f_{MB}(y)$,
and virtual baryons accompanying them,
$\langle n \rangle_{BM} = \int_{0}^{1} dy'\ f_{BM}(y')$:
\bg
\langle n \rangle_{MB} & = & \langle n \rangle_{BM}.
\label{qcon}
\en
This is just a statement of charge conservation.
Momentum conservation imposes the further requirement that
\bg
\langle y \rangle_{MB}\ +\ \langle y \rangle_{BM} & = & \langle n \rangle_{MB}
\label{mcon}
\en
where
$\langle y \rangle_{MB} = \int_{0}^{1} dy\ y\ f_{MB}(y)$ and
$\langle y \rangle_{BM} = \int_{0}^{1} dy'\ y'\ f_{BM}(y')$
are the average momentum fractions carried by meson $M$ and
the virtual baryon $B$, respectively.
Equations (\ref{qcon}) and (\ref{mcon}), and in fact similar relations
for all higher moments of $f(y)$, follow automatically from eqn.(\ref{pcon}).

In what follows we shall explicitly evaluate the functions $f_{MB}$ and
$f_{BM}$, and examine the conditions under which eqn.(\ref{pcon})
is satisfied.
The results will be used to calculate the contributions to the nucleon
structure function from the extended mesonic structure of the nucleon,
which are expressed as convolutions of the functions $f(y)$ with the
structure functions of the struck meson or baryon:
\bg
\delta^{(MB)} F_{2 N}(x)
& = & \int_{x}^{1} dy\ f_{MB}(y)\ F_{2 M}(x/y)
\label{delMF2} \\
\delta^{(BM)} F_{2 N}(x)
& = & \int_{x}^{1} dy'\ f_{BM}(y')\ F_{2 B}(x/y')
\label{delBF2}
\en
with $x = -q^{2}/2 p \cdot q$ being the Bjorken variable.
Note that eqns.(\ref{delMF2}) and (\ref{delBF2}) are correct when
physical (renormalised) meson---baryon coupling constants are used in the
functions $f_{MB}$ and $f_{BM}$ (see section 4 for details).
By comparing against the experimental structure functions, we will ultimately
test the reliability of the expansion in eqn.(\ref{state}), and in particular
the relative importance of the states involving vector mesons compared with
the pion states.

%%%%%%%%%%%%%%%%%%%%%%%%%%%%%%%%%%%%%%%%%%%%%%%%%%%%%%%%%%%%%%%%%%%%%%%%%%%%%%%
\section{The Pion-Nucleon Contribution}

%..............................................................................
\subsection{Covariant Formulation}

Traditionally the effects upon $F_{2N}(x)$ of the $\pi$
meson cloud have been studied most intensely.
The distribution function of a virtual pion accompanied by a recoiling
nucleon has been calculated in a covariant framework \cite{first,t83} as:
\bg
f_{\pi N}(y)
& = & \frac{3 g^{2}_{\pi N N}}{16 \pi^{2}}\
y\ \int_{-\infty}^{t^{N}_{max}} dt\
\frac{ (- t)\ {\cal F}_{\pi N}^{2}(t) }{ (t - m_{\pi}^{2})^{2} }.
\label{fpint}
\en
Here, $t \equiv k^{2} = t^{N}_{max} - k_{T}^{2}/(1-y)$ is the 4-momentum
squared of the virtual pion, with a kinematic maximum given by
$t^{N}_{max} = - m_{N}^{2} y^{2} / (1-y)$, and $k_{T}^{2}$ is the pion
transverse momentum squared.
In a covariant formulation the form factor, ${\cal F}_{\pi N}$,
parameterising the $\pi N N$ vertex, at which only the pion is
off-mass-shell, can only depend on $t$.

Contributions from processes in which the virtual nucleon (accompanied by
a recoiling pion) is struck have been calculated by several authors
\cite{hsb,zol,msm}, although not all agree.
Partly because there is less phenomenological experience with so-called
sideways form-factors (where the nucleon, rather than the pion, is
off-mass-shell) some early work \cite{bcw,sst,mts} simply defined
$f_{N\pi}(y')$ through eqn.(\ref{pcon}). However, this is unsatisfactory
from a theoretical point of view, and ideally we would like to verify
explicitly that the functions $f_{\pi N}$ and $f_{N\pi}$ satisfy
eqn.(\ref{pcon}).

Clearly the treatment of deep inelastic scattering from an interacting
nucleon is considerably more involved than that from a real nucleon,
which is described by the usual hadronic tensor
\bg
W^{\mu\nu}_{N}(p,q) & = & \gtuv\ W_{1N}(p,q)
                    \ + \ \tilde{p}^{\mu}\ \tilde{p}^{\nu}\ W_{2N}(p,q)
\label{wuvN}
\en
where $\gtuv = - g^{\mu\nu} + q^{\mu} q^{\nu} / q^{2}$ and
$\tilde{p}^{\mu} = p^{\mu} -  q^{\mu}\ p \cdot q / q^{2}$.
The hadronic vertex factor for the diagram of fig.1b in this case will be
\bg
{\rm Tr}\left[ (\not\!p + m_{N})\ i \gamma_{5}\ (\not\!p' + m_{N})\
               \hat{W}^{\mu\nu}_{N}(p',q)\
          (\not\!p' + m_{N})\ i \gamma_{5} \right]
\label{trace}
\en
where $\hat{W}^{\mu\nu}_{N}(p',q)$ is a matrix in Dirac space representing
the hadronic tensor for an interacting nucleon,
and is related to the hadronic tensor for real nucleons by \cite{bi}
\bg
W^{\mu\nu}_{N}(p,q)
& = &
\frac{1}{2} {\rm Tr}\left[
(\not\!p + m_{N})\ \hat{W}^{\mu\nu}_{N}(p,q) \right].
\label{wrel}
\en
If the struck nucleon is treated as an elementary fermion \cite{jaf} the
relevant operator in $\hat{W}^{\mu\nu}_{N}(p',q)$ is $\not\!q / 2 p' \cdot q$,
which leads to \cite{msm}
\bg
f_{N \pi}(y')
& = &
\frac{3 g^{2}_{\pi N N}}{16 \pi^{2}}\ y'\ \int_{-\infty}^{t'^{\pi}_{max}} dt'\
\left[ - m_{\pi}^{2} - \frac{1-y'}{y'} (t' - m_{N}^{2})  \right]
\frac{{\cal F}_{N \pi}^{2}(t')}{(t' - m_{N}^{2})^{2}},
\label{fnpit}
\en
where $t' \equiv p'^{2} = t'^{\pi}_{max} - p'^{2}_{T}/(1-y')$ is the
4-momentum squared of the virtual nucleon, with the upper limit now given by
$t'^{\pi}_{max} = m_{N}^{2} y' - m_{\pi}^{2} y' /(1-y')$, and $p'^{2}_{T}$
denotes the nucleon's transverse momentum squared.
Apart from the form factors, eqns.(\ref{fpint}) and (\ref{fnpit}) are
clearly related by an interchange $y' \leftrightarrow 1-y$.

Note that choosing a different operator form for $\hat{W}^{\mu\nu}_{N}$ may
lead to unphysical results. For example, with an operator involving $I$
rather than $\not\!\!q$ the trace factor in eqn.(\ref{fnpit}) is proportional
to $-m_{\pi}^{2}$.
Problems also arise for the emission of scalar or vector mesons \cite{scal}.
A full investigation of the off-mass-shell effects in deep inelastic structure
functions of composite objects will be the subject of a future publication
\cite{smt}.

The large-$t'$ suppression for the $N \pi N$ vertex is introduced
by the form factor ${\cal F}_{N\pi}$, which is usually parameterised by
a monopole or dipole function
\bg
{\cal F}_{N\pi}(t') & = &
\left( \frac{ \Lambda^{2}_{N\pi} - m_{N}^{2} }
            { \Lambda^{2}_{N\pi} - t' }
\right)^{n}
\nn
\en
for $n=1$ and 2, respectively.
However to satisfy eqn.(\ref{qcon}), the cut-off parameter $\Lambda_{N\pi}$
will in general have to be different from the cut-off $\Lambda_{\pi N}$
regulating the $\pi N N$ vertex form factor in eqn.(\ref{fpint}),
\bg
{\cal F}_{\pi N}(t) & = &
\left( \frac{ \Lambda^{2}_{\pi N} - m_{\pi}^{2} }
            { \Lambda^{2}_{\pi N} - t }
\right)^{n}.
\nn
\en
In general a different $\Lambda_{\pi N}$ would be required to satisfy
eqn.(\ref{mcon}), and it would not be possible to guarantee eqn.(\ref{pcon}).

An important assumption in the covariant convolution model is that the
dependence of the virtual meson and baryon structure functions in
eqns.(\ref{delMF2}) and (\ref{delBF2}) on the invariant mass squared
is negligible. The argument usually made is that the vertex form factor
suppresses contributions from the far off-mass-shell configurations
(ie. for $|t| \ageq 10\ m_{N}^{2}$ \cite{mts}).
However, in this approach even the identification of the off-shell structure
functions themselves is not very clear.
Some suggestions about how to relate the off-shell functions to the on-shell
ones were made \cite{nucl} in the context of DIS from nuclei, although
these were more {\em ad hoc} prescriptions rather than theoretical
derivations.
Attempts to simplify this situation were made in ref. \cite{inst}, where it
was proposed that the instant form of dynamics, where only on-mass-shell
particles are encountered, be used to calculate the nuclear
structure functions.
Along similar lines was the light-front approach of Berger {\em et al.}
\cite{bcw}.
Actually these two techniques are the same if one works in the infinite
momentum frame.
The instant form of dynamics was previously used by G\"{u}ttner {\em et al.}
\cite{gcpp} in the calculation of the function $f_{\pi N}(y)$ for the case of
pion electroproduction, and more recently by Zoller \cite{zol} in the
DIS of charged leptons from nucleons.

%..............................................................................
\subsection{Infinite Momentum Frame States}

An alternative to the use of covariant Feynman diagrams,
in the form of `old-fashioned' time-ordered perturbation theory in the
infinite momentum frame (IMF), was proposed some time ago by Weinberg
\cite{wein} for scalar particles. This was later extended by Drell, Levy
and Yan \cite{dly} to the $\pi N$ system in deep inelastic scattering.
The main virtues of this approach are that off-mass-shell ambiguities
in the structure functions of virtual particles can be avoided,
and that the meson and baryon distribution functions can be shown to
satisfy eqn.(\ref{pcon}) exactly.

In the time-ordered theory the analogue of fig.1a will now involve two
diagrams in which the $\pi$ moves forwards and backwards in time, fig.2.
However, in a frame of reference where the target nucleon is moving fast
along the $z$ direction with longitudinal momentum $p_{L} (\ra \infty)$, only
that diagram involving a forward moving pion gives a non-zero contribution.
In the IMF the target nucleon has energy
\bg
p_{0} & = & p_{L} + \frac{m_{N}^{2}}{2 p_{L}}
                  + \Ord.
\nn
\en
Following Weinberg \cite{wein} we write the pion 3-momentum as
\bg
{\bf k} & = & y\ {\bf p} + {\bf k}_{T},
\nn
\en
where ${\bf k}_{T} \cdot {\bf p} = 0$, and conservation of momentum
demands that the recoil nucleon momentum be
\bg
{\bf p'} & = & (1-y)\ {\bf p} - {\bf k}_{T}.
\nn
\en
Since all particles are on their mass shells the energies of
the intermediate $\pi$ and $N$ must be
\bg
k_{0} & = & |y|\ p_{L} + \frac{k_{T}^{2} + m_{\pi}^{2}}{2 |y| p_{L}}
                       + \Ord
\nn\\
p'_{0} & = & |1-y|\ p_{L} + \frac{k_{T}^{2} + m_{N}^{2}}{2 |1-y| p_{L}}
                          + \Ord.
\nn
\en
For forward moving particles (fig.2a) $y$ and $1-y$ are positive, and
according to the rules of the time-ordered perturbation theory the energy
denominator appearing in the calculation of $f_{\pi N}(y)$ is
$(p_{0} - p'_{0} - k_{0}) = (m_{N}^{2} - s_{\pi N}) / 2 p_{L}$, where
\bg
s_{\pi N} & = & s_{\pi N}(k^{2}_{T},y)
\ =\ (p'_{0} + k_{0})^{2} - ({\bf p'} + {\bf k})^{2}\
\ =\ \ \frac{ k^{2}_{T} + m_{\pi}^{2} }{ y }
     + \frac{ k^{2}_{T} + m_{N}^{2} }{ 1 - y }
\en
is the centre of mass energy squared of the intermediate $\pi N$ state.
Changing the variables of integration from $d^{3} {\bf k}$ to $d y$ and
$d k_{T}^{2}$, all powers of $p_{L}$ are seen to cancel when combined
with the appropriate vertex factors, $(2 p'_{0})^{-1}$ and $(2 k_{0})^{-2}$.
However, for a backward moving pion (fig.2b) $y$ is negative, and the energy
denominator becomes $(p_{0} - p'_{0} - k_{0}) = 2 y p_{L} + O(1/p_{L})$.
Therefore in the $p_{L} \ra \infty$ limit this time-ordering does not
contribute, and the result of eqn.(\ref{fpint}) is reproduced,
form factor aside.

For an interacting nucleon with $\pi$ recoil, fig.3, the kinematics are
similar to the above, namely the nucleon and pion move with 3-momenta
\bg
{\bf p'} & = & y'\ {\bf p} - {\bf k}_{T}
\nn\\
{\bf k} & = & (1-y')\ {\bf p} + {\bf k}_{T}
\nn
\en
and have energies
\bg
p'_{0} & = & |y'|\ p_{L} + \frac{k_{T}^{2} + m_{N}^{2}}{2 |y'| p_{L}}
                         + \Ord
\nn\\
k_{0} & = & |1-y'|\ p_{L} + \frac{k_{T}^{2} + m_{\pi}^{2}}{2 |1-y'| p_{L}}
                          + \Ord,
\nn
\en
respectively.
The general structure of the tensor describing a non-elementary interacting
nucleon can be written as
\bg
\hat{W}^{\mu\nu}_{N}(p,q)
& = & \gtuv\ \left( \hat{W}_{0}\
               \ +\ \not\!p\ \hat{W}_{1}\
               \ +\ \not\!q\ \hat{W}_{2} \right) \ \ +\ ...
\label{what}
\en
where we have omitted terms proportional only to $p^{\mu,\nu}$ and
$q^{\mu,\nu}$.
The functions $\hat{W}_{0,1,2}$ are related to the on-mass-shell structure
function $W_{1N}$ by eqn.(\ref{wrel}):
\bg
W_{1N}(p,q) & = & 2\ \left( m_{N} \hat{W}_{0}
                          + m_{N}^{2} \hat{W}_{1}
                          + p \cdot q\ \hat{W}_{2}
                     \right).
\en
Then direct evaluation of the trace in eqn.(\ref{trace}) gives
$$
4\ (2 p \cdot p'\ -\ 2 m_{N}^{2})
\left[ \gtuv\ \left( m_{N} \hat{W}_{0}
                \ +\ m_{N}^{2} \hat{W}_{1}
                \ +\ p' \cdot q\ \hat{W}_{2}
              \right)\ +\ ...
\right]
\ =\ 2\ (2 p \cdot p'\ -\ 2 m_{N}^{2})\ \gtuv\ W_{1N}(p',q)\ +\ ...
$$
where now the exact on-shell nucleon structure function
appears, and there is no off-shell ambiguity.

For a backward moving nucleon (fig.3b) $y'$ is negative, and
$2 p \cdot p' - 2 m_{N}^{2} = -4 y' p_{L}^{2} + O(1/p_{L})$, so that the
numerator becomes large in the $p_{L} \ra \infty$ limit.
Technically this is due to the `badness' of the operator $\gamma_{5}$,
which mixes upper and lower components of the nucleon spinors.
The energy denominator here is
$(p_{0} - p'_{0} - k_{0}) = 2 y' p_{L} + O(1/p_{L})$, and when squared and
combined with the $1/p_{L}^{2}$ from the integration and vertex factors,
the contribution from this diagram vanishes when $p_{L}$ is infinite.

Therefore we need only evaluate the diagram with the forward moving nucleon,
fig.3a, which gives the result of eqn.(\ref{fnpit}):
\bg
f_{N \pi}(y')
& = &
\frac{ 3 g^{2}_{\pi N N} }{ 16 \pi^{2} }\ \int_{0}^{\infty}\ dk_{T}^{2}\
\left[ \frac{ k_{T}^{2} + (1-y')^{2} m_{N}^{2} }
            { y' }
\right]
\frac{ {\cal F}_{N \pi}^{2}(k^{2}_{T},y') }
     { y' (1-y') (m_{N}^{2} - s_{N\pi})^{2} },
\label{fnpi}
\en
with $s_{N\pi}(k^{2}_{T},y') = s_{\pi N}(k^{2}_{T},1-y')$,
except that the form factor is now unknown.
It is quite natural to choose the form factor to be a function of the centre
of mass energy squared of the $\pi N$ system, $s_{N\pi}$, as was done by
Zoller \cite{zol}.
The only difference between our treatment and that in ref. \cite{zol} is that
we follow the conventional normalisation so that the coupling constant
$g_{\pi N N}$ has its standard value at the pole,
\bg
{\cal F}_{N \pi}(k^{2}_{T},y')
& = & \exp \left( \frac{ m_{N}^{2} - s_{N\pi} }{ \Lambda^{2} }
           \right).
\label{y'ff}
\en
Within this approach there is an explicit symmetry between the processes
in which the intermediate pion and the intermediate nucleon are struck,
provided we take the form factor in $f_{\pi N}$ as
\bg
{\cal F}_{\pi N}(k_{T}^{2},y) & = & {\cal F}_{N\pi}(k_{T}^{2},1-y).
\label{yff}
\en
Then as long as the same mass parameter $\Lambda$ is used in both vertex
functions, eqn.(\ref{pcon}) is automatically satisfied.

In fig.4 we compare $f_{\pi N}(y)$ with a dipole form factor and with
the form factor in eqn.(\ref{yff}).
In order to make the comparison meaningful the cut-offs have been chosen
to yield the same value of $\langle n \rangle_{\pi N}\ \ (\simeq 0.25)$.
With the $y$-dependent form factor in eqn.(\ref{yff})\ $f_{\pi N}(y)$ is a
little broader and peaks at around $y = 0.3$, compared with $y \simeq 0.2$
for the covariant formulation with a dipole form factor.
Consequently, the convolution of $f_{\pi N}(y)$ with $F_{2 M}$ for the
$y$-dependent form factor will have a slightly smaller peak and extend
to marginally larger $x$.
This is evident in fig.5, where we show the calculated SU(2) antiquark
contribution to $\delta^{(\pi N)} F_{2 p}(x)$, compared with some
recent empirical data at $Q^{2} = 4$ GeV$^{2}$.

%%%%%%%%%%%%%%%%%%%%%%%%%%%%%%%%%%%%%%%%%%%%%%%%%%%%%%%%%%%%%%%%%%%%%%%%%%%%%%%
\section{Vector meson content of the nucleon}

In this section we extend the convolution model analysis to the vector meson
sector.
Our approach is similar to that described in section 2.2, namely we use
time-ordered perturbation theory to evaluate those diagrams which are
non-zero in the IMF.
Previous calculations \cite{hsb,ss} of the vector meson contributions
were made in a covariant framework, but with the assumption that the
vector meson and nucleon intermediate states were on-mass-shell.
In our approach we self-consistently calculate both the contribution
from a struck vector meson (fig.6a) and from a struck nucleon with
a vector meson recoil (fig.6b),
and show explicitly that the distribution functions for these obey the
relation in eqn.(\ref{pcon}) exactly.

Starting from the effective $V N N$ Lagrangian (see e.g. ref. \cite{bonn}),
where $V = \rho$ or $\omega$, we write in full the vector meson contribution
(with a nucleon recoil) to the nucleon hadronic tensor:
\bg
\delta^{(VN)} W^{\mu\nu}(p,q)
& = & c_{V} \int \frac{ d^{3}{\bf k} }{ (2\pi)^{3} (2 p'_{0}) (2 k_{0})^{2} }
\left( g_{\vn}^{2}\ A_{\alpha\beta}
      \ +\ \frac{ f_{\vn}^{2} }{ (4 m_{N})^{2} }\ B_{\alpha\beta}
      \ +\ g_{\vn} \frac{ f_{\vn} }{ 4 m_{N} }\ C_{\alpha\beta}
\right)
\nn\\
&   &
\hspace*{1cm}
\cdot \frac{ {\cal F}_{VN}^{2}(k_{T}^{2},y) }
           { (p_{0} - p'_{0} - k_{0} )^{2} }\
W^{\mu\nu\alpha\beta}_{V}(k,q),
\label{delWuv}
\en
where
\bg
A_{\alpha\beta} & = & 2\ (m_{N}^{2} - p \cdot p')\ g_{\alpha\beta}\
                   +\ 2 p_{\alpha} p'_{\beta}\
                   +\ 2 p'_{\alpha} p_{\beta}\
\label{vnn}
\nn\\
B_{\alpha\beta} & = & \frac{1}{2}
\left[
(m_{V}^{2}\ m_{N}^{2} - 2\ p \cdot k\ p' \cdot k\
+\ m_{V}^{2} p \cdot p')\ g_{\alpha\beta}\
-\ (m_{N}^{2} + p \cdot p')\ k_{\alpha} k_{\beta}
\right.
\nn\\
&   &
\left.
-\ m_{V}^{2}\ (p_{\alpha} p'_{\beta} + p_{\beta} p'_{\alpha})\
+\ p' \cdot k\ (p_{\alpha} k_{\beta} + p_{\beta} k_{\alpha})\
+\ p \cdot k\ (p'_{\alpha} k_{\beta} + p'_{\beta} k_{\alpha})\
\right]
\\
C_{\alpha\beta} & = &
2\ (p \cdot k - p' \cdot k)\ g_{\alpha\beta}\
-\ (p_{\alpha} k_{\beta} + p_{\beta} k_{\alpha})\
+\ (p'_{\alpha} k_{\beta} + p'_{\beta} k_{\alpha})\
\nn
\en
are the $V N N$ vertex trace factors for the vector, tensor
and vector-tensor interference couplings, respectively.
The isospin factor $c_{V}$ is equal to 3 and 1 for isovector
and isoscalar mesons, respectively.
For an on-mass-shell vector meson, the spin-1 tensor $W^{\mu\nu\alpha\beta}$,
symmetric under the interchange of $\mu \leftrightarrow \nu$ and
$\alpha \leftrightarrow \beta$, is given by:
\bg
W^{\mu\nu\alpha\beta}(k,q)
& = & \left( \gtuv\ W_{1V}(k,q)\ +\ \ktu\ \ktv\ W_{2V}(k,q)
      \right)\ \gtab.
\en
This form guarantees that the vector current is conserved,
$k_{\alpha,\beta} W^{\mu\nu\alpha\beta} = 0
= q_{\mu,\nu} W^{\mu\nu\alpha\beta}$.
Furthermore, it reproduces the correct unpolarised on-shell spin 1 tensor
when contracted with the meson polarisation vectors
($\epsilon_{\alpha,\beta}$) and summed over the $V$ helicity, $\lambda$
\cite{s1on}:
\bg
W^{\mu\nu}_{V}(k,q)
& = & \sum_{\lambda} \epsilon_{\alpha}^{*}(\lambda,k)\
                     \epsilon_{\beta}(\lambda,k)\
               W^{\mu\nu\alpha\beta}_{V}(k,q)
\nn\\
& = & \left( - g_{\alpha\beta} + \frac{k_{\alpha} k_{\beta}}{k^{2}} \right)
               W^{\mu\nu\alpha\beta}_{V}(k,q)
\\
& \propto & \gtuv\ W_{1V}(k,q)\ +\ \ktu\ \ktv\ W_{2V}(k,q).
\nn
\en
In the case of DIS from a vector particle emitted by a nucleon, fig.6a,
contracting the spin 1 tensor $W^{\mu\nu\alpha\beta}$ with the $V N N$
vertex trace factors in eqn.(\ref{vnn}), and equating coefficients of $\gtuv$
gives:
\bg
\delta^{(VN)} W_{1N}(p,q)
& = & c_{V} \int \frac{ d^{3}{\bf k} }{ (2\pi)^{3} (2 p'_{0}) (2 k_{0})^{2} }
\left\{
g_{\vn}^{2} \left[ - 6 m_{N}^{2} + \frac{4 p \cdot k\ p' \cdot k}{m_{V}^{2}}
                                 + 2 p \cdot p'
            \right]
\right.
\nn\\
&   & \hspace*{0.5cm}
\ -\ \frac{ f_{\vn}^{2} }{ 2 }\
     \left[ - 3 m_{V}^{2} + \frac{ 4 p \cdot k\ p' \cdot k }{ m_{N}^{2} }
                          - \frac{ m_{V}^{2} p \cdot p' }{ m_{N}^{2} }
     \right]
\nn\\
&   & \hspace*{0.5cm}
\left.
\ -\ 6\ g_{\vn}\ f_{\vn}\  \left[ p \cdot k - p' \cdot k \right]
\right\}
\frac{ {\cal F}_{VN}(k^{2}_{T},y) }
     { (m_{N}^{2} - s_{VN})^{2} } W_{1V}(k,q).
\en
Using the IMF kinematics (which are similar to those for the $\pi N$
system, except that $m_{\pi} \ra m_{V}$) together with the Callan-Gross
relation for the nucleon and vector meson, enables the contribution to
$F_{2N}$ from vector mesons to be written as a convolution of the vector
meson distribution function $f_{VN}(y)$ with the on-shell vector meson
structure function $F_{2V}(x/y)$, as in eqn.(\ref{delMF2}), where now
\bg
\hspace*{-1cm}
f_{VN}(y)
& = & \frac{ c_{V} }{ 16 \pi^{2} }\ \int_{0}^{\infty} dk_{T}^{2}
\left\{
g_{\vn}^{2}
\left[ \frac{ \left( k_{T}^{2} + y^{2} m_{N}^{2} + m_{V}^{2} \right)
            \left( k_{T}^{2} + y^{2} m_{N}^{2} + (1-y)^{2} m_{V}^{2} \right) }
            { y^{2} (1-y) m_{V}^{2} }
     + \frac{ k_{T}^{2} + y^{2} m_{N}^{2} }{ 1-y }
     - 4 m_{N}^{2}
\right]
\right.
\nn\\
&   &
\ +\ f_{\vn}^{2}\
\left[ \frac{ \left( k_{T}^{2} + y^{2} m_{N}^{2} + m_{V}^{2} \right)
            \left( k_{T}^{2} + y^{2} m_{N}^{2} + (1-y)^{2} m_{V}^{2} \right) }
            { 2 y^{2} (1-y) m_{N}^{2} }
     - \frac{ m_{V}^{2} \left( k_{T}^{2} + (2-y)^{2} m_{N}^{2} \right) }
            { 4 (1-y) m_{N}^{2} }
     - m_{V}^{2}
\right]
\nn\\
&   &
\left.
\ +\ 3\ g_{\vn}\ f_{\vn}\
\left[ \frac{ k_{T}^{2} + y^{2} m_{N}^{2} - (1-y)\ m_{V}^{2} }{ 1-y }
\right]
\right\}
\frac{ {\cal F}_{VN}^{2}(k_{T}^{2},y) }
     { y (1-y) (m_{N}^{2} - s_{VN})^{2} }.
\label{fvn}
\en
The $VNN$ form factor is defined analogously to eqn.(\ref{yff}),
\bg
{\cal F}_{VN}(k_{T}^{2},y)
& = & \exp \left( \frac{ m_{N}^{2} - s_{VN} }{ \Lambda^{2} }
           \right)
\label{vff}
\en
and the $V N$ centre of mass energy squared is
\bg
s_{VN} & = & s_{VN}(k_{T}^{2},y)
\ =\ \frac{ k_{T}^{2} + m_{V}^{2} }{ y }
\ +\ \frac{ k_{T}^{2} + m_{N}^{2} }{ 1 - y }.
\en
Suppression of backward moving vector mesons is achieved in the IMF
by the energy denominators, as for pions.
The vector meson structure function $F_{2 V}$
is not known experimentally, so in our numerical calculations we assume
that its $x$-dependence resembles that of the $\pi$ meson
structure function, which has been determined experimentally \cite{na10}.

For the vector meson recoil process, fig.6b, we evaluate the distribution
function $f_{NV}(y')$ using the full spinor structure of
$\hat{W}^{\mu\nu}_{N}$ in eqn.(\ref{what}):
\bg
\delta^{(NV)} W_{1N}(p,q)
& = & c_{V}
\int \frac{ d^{3}{\bf p'} }{ (2\pi)^{3} (2 p'_{0})^{2} (2 k_{0}) }
\left( g_{\vn}^{2}\ A_{\alpha\beta}
      \ +\ \frac{ f_{\vn}^{2} }{ (4 m_{N})^{2} }\ B_{\alpha\beta}
      \ +\ g_{\vn} \frac{ f_{\vn} }{ 4 m_{N} }\ C_{\alpha\beta}
\right)
\nn\\
&   & \hspace*{-1.5cm}
\cdot \sum_{\lambda}\ \epsilon^{\ast}_{\alpha}(\lambda,k)\
                      \epsilon_{\beta}(\lambda,k)\
\frac{ {\cal F}_{NV}^{2}(k^{2}_{T},y') }
     { (p_{0} - p'_{0} - k_{0} )^{2} }\
\left( 2 m_{N}\ \hat{W}_{0} + 2 m_{N}^{2}\ \hat{W}_{1}
                            + 2 p' \cdot q\ \hat{W}_{2} \right)
\label{wnv}
\en
where the tensors $A, B$ and $C$ are as in eqn.(\ref{vnn}).
Performing the contractions over the indices $\alpha, \beta$ leads
to the convolution integral of eqn.(\ref{delBF2}), with the nucleon
distribution function with a vector meson recoil given by
\bg
\hspace*{-1cm}
f_{NV}(y')
& = & \frac{ c_{V} }{ 16 \pi^{2} }\ \int_{0}^{\infty} dk^{2}_{T}
\left\{
g_{\vn}^{2}
\left[ \frac{ \left( k^{2}_{T} + (1-y')^{2} m_{N}^{2} + m_{V}^{2} \right)
            \left( k^{2}_{T} + (1-y')^{2} m_{N}^{2} + y'^{2} m_{V}^{2}\right)}
            { y' (1-y')^{2} m_{V}^{2} }
\right.
\right.
\nn\\
&   & \hspace*{4cm}
\left.
    \ +\ \frac{ k^{2}_{T} + (1-y')^{2} m_{N}^{2} }{ y' }
    \ -\ 4 m_{N}^{2}
\right]
\nn\\
&   & \hspace*{-1cm}
\ +\ f_{\vn}^{2}\
\left[ \frac{ \left( k^{2}_{T} + (1-y')^{2} m_{N}^{2} + m_{V}^{2} \right)
            \left( k^{2}_{T} + (1-y')^{2} m_{N}^{2} + y'^{2} m_{V}^{2}\right)}
            { 2\ y'\ (1-y')^{2}\ m_{N}^{2} }
     - \frac{ m_{V}^{2} (k^{2}_{T} + (1+y')^{2} m_{N}^{2}) }
            { 4\ y'\ m_{N}^{2} }
     - m_{V}^{2}
\right]
\nn\\
&   & \hspace*{-1cm}
\left.
\ +\ 3\ g_{\vn}\ f_{\vn}\
\left[ \frac{ k^{2}_{T} + (1-y')^{2} m_{N}^{2} - y' m_{V}^{2} }{ 1-y' }
\right]
\right\}
\frac{ {\cal F}_{NV}^{2}(k^{2}_{T},y') }
     { y' (1-y') (m_{N}^{2} - s_{NV})^{2} }.
\label{fnv}
\en
and where $s_{NV}(k^{2}_{T},y') = s_{VN}(k^{2}_{T},1-y')$.
Again, we have evaluated only the diagram with forward moving nucleons
which is non-zero in the IMF.
It is clear therefore from eqns.(\ref{fvn}) and (\ref{fnv})
that the probability distributions for the $V N$ intermediate states
are related by $f_{NV}(y') = f_{VN}(1-y')$.

Our numerical results, which are discussed below, rely upon the physical
vector meson---nucleon coupling constants whose values are taken at the poles,
as obtained from analyses of $\pi N$ scattering data:
$g_{\rho N N}^{2} / 4\pi = 0.55$,
$f_{\rho N N} / g_{\rho N N} = 6.1$ \cite{hp},
and
$g_{\omega N N}^{2} / 4\pi = 8.1$,
$f_{\omega N N} / g_{\omega N N} = 0$ \cite{gk}.

%%%%%%%%%%%%%%%%%%%%%%%%%%%%%%%%%%%%%%%%%%%%%%%%%%%%%%%%%%%%%%%%%%%%%%%%%%%%%%%
\section{Results and Discussion}

Fig.7 shows the meson distribution functions $f_{\rho N}$, $f_{\omega N}$ and
$f_{\pi N}$ (scaled by a factor 1/3) for the same vertex cut-off parameter
$\Lambda (=1.4$ GeV).
The vector meson component will only be relevant when very hard form factors
are employed.
To make this point more explicit, we plot in fig.8 the average multiplicities
$\langle n \rangle_{V N}$ and $\langle n \rangle_{\pi N}$ as a function
of $\Lambda$.
The dependence on $\Lambda$ is much stronger for the $\rho$ than for
$\pi$ mesons.
For $\Lambda \aleq 1.4$ GeV, $\langle n \rangle_{\rho N}$ is considerably
smaller than $\langle n \rangle_{\pi N}$, and it is only with much larger
cut-offs ($\Lambda \ageq 1.8$ GeV) that the $\rho$ multiplicity becomes
comparable with that of the $\pi$.
Note that $\Lambda = (1000, 1400, 1800)$ MeV corresponds to a dipole
$\Lambda_{\pi N} \simeq (650, 1020, 1410)$ MeV for the same
$\langle n \rangle_{\pi N}$.

One should observe that the trace factor inside the braces in $f_{V N}(y)$
is divergent in the limit $y \ra 0$, so that use of a form factor
$\propto \exp \left[ y\ (m_{N}^{2} - s_{VN}) \right]$,
which corresponds to a $t$-dependent covariant form factor
$\exp \left[ t - m_{V}^{2} \right]$,
would make $\delta^{(VN)} F_{2N}(x)$ approach a finite value
as $x \ra 0$, much like for a perturbative sea distribution.
However, there are several problems with accepting such a result, the most
obvious of which is that it would violate charge and momentum conservation
very badly, since $f_{N V}(y') \ra 0$ for $y' \ra 1$ and $\ra$ constant as
$y' \ra 0$ for a form factor
$\propto \exp \left[ y' (m_{N}^{2} - s_{NV}) \right]$,
which in the covariant formalism corresponds to\ \
$\exp \left[ t' - m_{N}^{2} \right]$.
Furthermore, it would lead to a gross violation of the Adler sum rule, which
integrates the flavour combination $u - \bar{u} - d + \bar{d}$,
and such a violation has not been observed in the range
$1 < Q^{2} < 40$\ GeV$^{2}$ \cite{adl}.
This gives further evidence for the preference of the IMF approach together
with the form factor in eqn.(\ref{vff}).
Note, however, that because the baryon recoil contributions to the quark and
antiquark distributions are related by
\bg
\delta^{(MB)} u(x) = \delta^{(MB)} \bar{d}(x), \ \ \ \ \ \
\delta^{(MB)} d(x) = \delta^{(MB)} \bar{u}(x)
\label{delrel}
\en
the divergent contributions would cancel for
the Gottfried (which depends on the combination $u + \bar{u} - d - \bar{d}$)
and Gross--Llewellyn-Smith ($u - \bar{u} + d - \bar{d}$) sum rules.

In previous studies \cite{t83,fms} restrictions have been obtained
on the magnitude of the form factor cut-offs by comparing
$\langle y \rangle_{MB}$ with the measured momentum fractions carried by
the antiquarks. Even more stringent constraints can be achieved by also
demanding that the shape of the meson exchange contributions to $\bar{q}(x)$,
\bg
\delta^{(MB)} \bar{q}(x)
& = & \int_{x}^{1} \frac{ dy }{ y }\ f_{MB}(y)\ \bar{q}_{M}(x/y)
\label{delq}
\en
be consistent with the shape of the experimental antiquark distribution
\cite{fms,kl}.
Fig.9 shows the calculated antiquark distributions from the $\pi$ component
of the nucleon alone and from the pion plus vector meson structure of the
nucleon, for $\Lambda = 1.2$ and 1.4 GeV.
Clearly the SU(2)\ $\bar{q}$ content of the nucleon (as parameterised by Owens,
Morfin and Tung, Eichten {\em et al.} and Diemoz {\em et al.} \cite{param})
is saturated for $\Lambda \approx 1.2$ GeV in the intermediate-$x$ region.
For the $\pi N N$ vertex this corresponds to a dipole form factor cut-off
$\Lambda_{\pi N} \approx 830$ MeV --- considerably smaller than that used
by many authors.
We can conclude therefore that for the range of form factor cut-offs
allowed by the data, vector mesons play only a marginal role in the DIS
process.
The maximum value of $\Lambda$ would have to be even smaller with the
inclusion of $\pi \Delta$ states in the nucleon, as it has been shown
previously \cite{sst,kl,mts,hsb} that these give non-negligible contributions
to the nucleon structure function.
(We have not included the $\rho \Delta$ states as these will be insignificant
for the range of $\Lambda$ considered here.)
The $\pi \Delta$ states would also be of relevance to the calculated
$\bar{d}-\bar{u}$ difference (and to the Gottfried sum rule) resulting
from DIS from the $\pi N$ and $\rho N$ components,
which will be partly cancelled by this contribution.

At this point we would like to clarify an issue that has been the cause
of some confusion recently in the literature.
The meson and baryon exchange diagrams in fig.1 describe physical processes
(inclusive baryon and meson leptoproduction) whose cross sections involve
physical (renormalised) coupling constants.
When integrated over the recoil particles' momenta these yield the
inclusive DIS cross sections, which are proportional to the total quark
(and antiquark) distributions
\bg
q(x)
& = & Z\ q_{\rm bare}(x)
\ +\  \sum_{MB} \left( \delta^{(MB)} q(x)\ +\ \delta^{(BM)} q(x) \right).
\label{qtilde}
\en
Therefore $\delta q(x)$, and the convolution integrals in eqns.(\ref{delMF2}),
(\ref{delBF2}) and (\ref{delq}), are expressed in terms of renormalised
coupling constants contained in the functions $f(y)$.
{}From eqn.(\ref{qtilde}) we also determine the bare nucleon probability
\bg
Z & = & 1\ -\ \sum_{M B} \langle n \rangle_{MB}
\label{z}
\en
by demanding that the valence number and momentum sum rules are satisfied.
We emphasise that all quantities in eqns.(\ref{qtilde}) and (\ref{z}) are
evaluated using renormalised coupling constants.

We could, of course, choose to work at a given order in the bare coupling
constant, and explicitly verify that the various sum rules are satisfied.
For example, to lowest order ($g_{0}^{2}$) the total quark distributions
would be \cite{smst}
\bg
q(x)
& = & Z\ \left\{
      q_{\rm bare}(x)
\ + \ \sum_{MB} \left( \delta^{(MB)} q_{(0)}(x)\ +\ \delta^{(BM)} q_{(0)}(x)
                \right)
         \right\}
\hspace*{1cm} ({\rm order}\ \ g_{0}^{2})
\label{qtilde2}
\en
with
\bg
Z & = & \left( 1 + \sum_{M B} \langle n_{(0)} \rangle_{MB} \right)^{-1}
\hspace*{2cm} ({\rm order}\ \ g_{0}^{2}),
\label{z2}
\en
where the subscript $(0)$ indicates that the functions $f(y)$ here are
evaluated using bare couplings.
Eqns.(\ref{qtilde}) and (\ref{z}) are easily recovered since the bare
couplings, to this order, are defined by $g_{0}^{2} = g_{{\rm ren}}^{2} / Z$.
It would, however, be inconsistent to use eqns.(\ref{qtilde2}) and (\ref{z2})
with renormalised coupling constants, especially with large form factor
cut-offs.
As long as the form factors are soft, the difference between
the bare and renormalised couplings is quite small.
However, with large cut-off masses the bare couplings would need to be
substantially bigger than the physical ones.
(In fact, the form factor cut-off dependence of the bare $\pi N$ coupling
constant in the cloudy bag model \cite{tmt} showed some 40\% difference
for very hard form factors --- or small bag radii, $\sim 0.6$ fm.)
In addition, with large values of $\Lambda$ the higher order diagrams
involving more than one meson in the intermediate state would become
non-negligible, and the initial assumption that the series in
eqn.(\ref{state}) can be truncated at the one-meson level would be
seriously in doubt.
Fortunately, we need not consider the multi--meson contributions,
since fig.9 clearly demonstrates
the difficulty in reconciling the empirical data with quark distributions
calculated with such large cut-offs.

Finally, some additional comments regarding the justification
of our calculation in terms of an incoherent summation of cross sections
for the various meson exchange processes.
Because of the pseudoscalar (or pseudovector) nature of the $\pi N N$ vertex,
there is no interference between $\pi$ meson and vector meson exchange.
Furthermore, there will be no mixing between the $\omega$ and $\rho$ exchange
configurations due to their different isospins.
In fact, all of the processes considered in this analysis can be added
incoherently.
The question remains, however, whether it will be possible to identify
an explicit vector meson contribution to $F_{2N}(x)$ in an unambiguous way
in deep inelastic scattering experiments.
While it may be feasible to search for one pion exchange by observing
the distribution of the produced low-momentum baryon spectrum \cite{semi},
because of the smaller absolute vector meson cross section it will be
difficult to separate this component from both the perturbative background
and from that due to other mesons.

\vspace*{1cm}

%%%%%%%%%%%%%%%%%%%%%%%%%%%%%%%%%%%%%%%%%%%%%%%%%%%%%%%%%%%%%%%%%%%%%%%%%%%%%%%
\hspace*{-0.5cm}
{\large {\bf Acknowledgements}}

\hspace*{-0.5cm}
We would like to thank E.M.Henley for his friendly assistance and
continued interest in this problem, as well as S.D.Bass, C.A.Hurst,
W.-Y.P.Hwang, N.N.Nikolaev, A.W.Schreiber, J.Speth and V.R.Zoller
for useful comments and discussions related to this work.
W.M. would like to thank the Institute for Nuclear Theory at the
University of Washington for its hospitality during a recent visit,
where part of this work was carried out.
This work was supported by the Australian Research Council.

\newpage

%%%%%%%%%%%%%%%%%%%%%%%%%%%%%%%%%%%%%%%%%%%%%%%%%%%%%%%%%%%%%%%%%%%%%%%%%%%%%%%

\newpage
\vspace*{2cm}

\hspace*{-0.5cm}
{\large {\bf Figure captions}}

\vspace*{0.5cm}

\normalsize

{\bf 1.} Deep inelastic scattering from the virtual\ (a) meson and\ (b) baryon
         components of a physical nucleon.

{\bf 2.} Time-ordered diagrams for pions moving\ (a) forwards
         and\ (b) backwards in time.
         Time is increasing from left to right.

{\bf 3.} Time-ordered diagrams for nucleons moving\ (a) forwards
         and\ (b) backwards in time.

{\bf 4.} $\pi N$ distribution function for a dipole form factor and that given
         in eqn.(\ref{yff}). The cut-offs are chosen so that
         $\langle n \rangle_{\pi N} \simeq 0.25$ in both cases.

{\bf 5.} Proton SU(2) antiquark distributions from DIS on the $\pi N$
         component of the nucleon, evaluated for the different $\pi N$
         form factors, as in fig.4.
         The data (dotted curves) are the parameterisations of Owens,
         Morfin and Tung, Eichten {\em et al.} and Diemoz {\em et al.}
         \cite{param}.

{\bf 6.} Time-ordered diagrams for the DIS from\ (a) vector mesons and\
         (b) nucleons with recoil vector mesons,
         that are non-zero in the IMF.

{\bf 7.} Meson distribution functions $f_{\rho N}(y)$, $f_{\omega N}(y)$
         and $f_{\pi N}(y)$, for $\Lambda = 1.4$GeV.
         Note the pion distribution is scaled by a factor of 1/3.

{\bf 8.} Average number densities for the $\pi$, $\rho$ and $\omega$
         mesons in a nucleon, as a function of the meson-nucleon
         form factor cut-off.

{\bf 9.} Proton SU(2) antiquark distributions, calculated with $\pi$ and
         $\pi + \rho + \omega$ components in the nucleon.
         The lower (upper) solid and dashed curves correspond to
         $\Lambda = 1.2$ (1.4) GeV.
         The data are from ref. \cite{param}.


\begin{thebibliography}{99}
\bibitem{scal}  For scalar mesons, the trace factor in $f_{N \sigma}(y')$\
                for the structure\ $\not\!q / 2 p' \cdot q$\ is\
                $4\ m_{N}^{2}\ -\ m_{\sigma}^{2}\\
                             \ +\ (m_{N}^{2} - t')\ (1-y')/y' $,
                which is clearly related to the trace in $f_{\sigma N}(y)$,
                namely $-t + 4 m_{N}^{2}$ when written in terms of the
                transverse momentum squared (with $m_{\pi} \ra m_{\sigma}$),
                and which thereby satisfies eqn.(\ref{pcon}), apart from
                the form factor.
                For an operator $I/(2 m_{N})$ the trace factor in
                $f_{N \sigma}(y')$ is $2 t' + 2 m_{N}^{2} - m_{\sigma}^{2}$,
                but this not only violates baryon number conservation,
                it also leads to an unphysical (negative) cross section.
                Similar problems appear in $f_{N V}(y')$ with such an
                operator.
\end{thebibliography}
\end{document}